\documentstyle[12pt]{article}
\begin{document}
\begin{flushright}
NUP-A-96-6\\
June, 1996
\end{flushright}
\hspace{\fill}
\vspace{14mm}
\renewcommand{\thefootnote}{\fnsymbol{footnote}}
\begin{center}
\Large\bf{A Yang--Mills Theory in Loop Space\\ 
and Chapline--Manton Coupling}
\end{center}

\hspace{\fill}

\begin{center}
\large{Shinichi DEGUCHI
\footnote{E-mail: deguchi@phys.cst.nihon-u.ac.jp} 
and Tadahito NAKAJIMA
\footnote{E-mail: nakajima@phys.cst.nihon-u.ac.jp}}
\end{center}
\begin{center}
\it{Department of Physics and Atomic Energy Research Institute, \\
 College of Science and Technology, Nihon University, \\
 Kanda-Surugadai, Chiyoda-ku, Tokyo 101, Japan} 
\end{center}
\hspace{\fill}
\vspace{5mm}
\begin{center}
\bf{Abstract}
\end{center}

We consider a Yang--Mills theory in loop space whose gauge group is a  
Kac--Moody group with the central extension. From this theory, we derive 
a local field theory constructed of Yang--Mills fields and abelian 
antisymmetric and symmetric tensor fields of the second rank. 
The Chapline--Manton coupling, that is, coupling of Yang--Mills fields and 
a second-rank antisymmetric tensor field via the Chern--Simons 3-form is 
obtained in a systematic manner.  
\newpage
\renewcommand{\thefootnote}{\arabic{footnote}}
\setcounter{footnote}{0}

\section{Introduction}

Recently two types of gauge theories defined in loop space (the space of 
all loops in space--time) have been considered. One is a U(1) gauge theory in 
loop space [1,2]. It was shown that this yields a local field theory of the 
second-rank antisymmetric tensor field and  the Stueckelberg formalism for 
massive vector and massive third-rank tensor fields.  
The other is a Yang--Mills theory in loop space whose gauge group 
is a Kac--Moody group without the central extension [3]. From this theory,  
a non-abelian Stueckelberg formalism for massive second-rank tensor fields 
was derived in addition to the local Yang--Mills theory. 
All the local fields treated in the Yang--Mills and the U(1) gauge theories 
in loop space are geometrically characterized as constrained connection  
1-forms on the loop space.   
It should be noted that the U(1) gauge theory in loop space is different  
from the Yang--Mills theory in loop space with the U(1) Kac--Moody gauge group.

A fundamental property of Kac--Moody groups is the existence of central  
extensions [4].
In the present paper we construct a Yang--Mills theory in loop space whose 
gauge group is a Kac--Moody group {\it with the central extension}, 
which we call the {\it extended} Yang--Mills theory (EYMT) in loop space.  
This theory can be understood as a unified theory of the Yang--Mills 
and the U(1) gauge theories in loop space.  
From the EYMT in loop space, we derive a local field theory constructed of 
Yang--Mills fields and abelian antisymmetric and symmetric tensor fields of 
the second rank. 
In this case, it is remarkable that the EYMT in loop space yields the 
Chapline--Manton coupling, that is, coupling of Yang--Mills fields and  
a second-rank antisymmetric tensor field via the Chern--Simons 3-form [5]. 
Chapline and Manton found this coupling in the study of a unification of $N=1$ 
supergravity and $N=1$ supersymmetric Yang--Mills theory in 10 dimensions. 
The Chapline--Manton coupling also occurs in the low-energy limit of type 
I superstring theories [6]. 
As will be seen in the present paper, we can naturally obtain the 
Chapline--Manton coupling within the framework of Yang--Mills theories.    
This is an attractive property of the EYMT in loop space. 

\vspace{1mm}

\section{Brief Review}

\noindent
{\it 2.1. A U(1) gauge theory in loop space} 

\vspace{3mm}

We define a loop space $\Omega M^{D}$ as the set of all loops  
in $D$-dimensional Minkowski space $M^{D}$.  
An arbitrary loop $x^{\mu}=x^{\mu}(\sigma)$ [$0\leq\sigma\leq2\pi$,  
$x^{\mu}(0)=x^{\mu}(2\pi)$] in $M^{D}$ is represented as a point in  
$\Omega M^{D}$ denoted by coordinates $(x^{\mu\sigma})$ with $x^{\mu\sigma}
\equiv x^{\mu}(\sigma)$$\,$
\footnote{$^{)}$ In this paper, the indices $\kappa$, $\lambda$, $\mu$ and 
$\nu$ take the values 0, 1, 2, ..., $D-1$,  
while the indices $\rho$, $\sigma$, $\chi$ and $\omega$ take continuous  
values  from 0 to 2${\pi}$.}$^{)}$. 

Let us first review a U(1) gauge theory in the loop space $\Omega M^{D}$ 
[1,2]. The infinitesimal gauge transformation of 
a U(1) gauge field ${\cal A}^{U}_{\mu\sigma}[x] $ on $\Omega M^{D}$ 
is given by
\begin{eqnarray}
\delta{\cal A}^{U}_{\mu\sigma}[x]=
\partial_{\mu\sigma} \Lambda^{U}[x] \; ,  
\end{eqnarray}
%
where $\partial_{\mu\sigma} \equiv \partial/\partial x^{\mu\sigma}$, and 
$\Lambda^{U}$ is an infinitesimal scalar function on $\Omega M^D$.
Combining (1) and the reparametrization-invariant condition for 
$\Lambda^{U}$, 
\begin{eqnarray}
x'^{\mu}(\sigma)\partial_{\mu\sigma}\Lambda^{U}[x]=0  \;,  
\end{eqnarray}
%
we have  
\begin{eqnarray}
x'^{\mu}(\sigma){\cal A}^{U}_{\mu\sigma}[x]=0  \;,
\end{eqnarray}
%
where the prime indicates differentiation with respect to $\sigma$.

The simplest solution of (2) is 
\begin{eqnarray}
\Lambda^{U(0)}[x]\equiv
\int_{0}^{2\pi} {{d\sigma}\over{2\pi}} q x'^{\mu}(\sigma)
\lambda_{\mu}(x(\sigma)) \;,   
\end{eqnarray}
%
where $q$ is a constant with dimensions of ${\rm [length]}^{-1}$   
and $\lambda_{\mu}$ is an infinitesimal vector function on $M^{D}$. 
Corresponding to (4), we take  
\begin{eqnarray}
{\cal A}^{U(0)}_{\mu\sigma}[x]\equiv
q x'^{\nu}(\sigma)B_{\mu\nu}(x(\sigma)) \; 
\end{eqnarray}
%
as ${\cal A}^{U}_{\mu\sigma}$,  
where $B_{\mu\nu}$ is a local antisymmetric tensor field on $M^{D}$. 
Obviously ${\cal A}^{U(0)}_{\mu\sigma}$ satisfies (3). 
Substituting (4) and (5) into (1), we have the well-known gauge 
transformation 
$\delta B_{\mu\nu}=\partial_{\mu} \lambda_{\nu}
-\partial_{\nu} \lambda_{\mu}$.  
The field strength of ${\cal A}^{U(0)}_{\mu\sigma}$ is written in terms of 
that of $B_{\mu\nu}$. 
Using this result, we can show that the Maxwell action for  
${\cal A}^{U(0)}_{\mu\sigma}$ becomes the Kalb--Ramond action for  
$B_{\mu\nu}$. The local field theory of $B_{\mu\nu}$ is thus derived from  
the U(1) gauge theory in loop space.

In addition to ${\cal A}^{U(0)}_{\mu\sigma}$, 
two solutions of (3) has been examined in detail. 
For the solution consisting of vector and scalar fields on $M^{D}$, the U(1) 
gauge theory in loop space yields the Stueckelberg formalism for a massive 
vector field [1], while for the solution consisting of third-rank and  
second-rank tensor fields on $M^{D}$, it yields the Stueckelberg formalism  
extended to a third-rank tensor field [2]. 

\vspace{3mm}

\noindent
{\it 2.2. A Yang--Mills theory in loop space} 

\vspace{2.5mm}

Next we review a Yang--Mills theory in the loop space $\Omega M^{D}$ [3].  
We assume that the gauge group is a Kac--Moody group $\hat{G}_{0}$ whose 
generators $T_{a}(\sigma)$ satisfy the commutation relations 
\begin{eqnarray}
[^{\,} T_{a}(\rho),\, T_{b}(\sigma)]=
i{f_{ab}}^{c}\delta(\rho-\sigma)T_{c}(\rho) \;,
\end{eqnarray}
%
where ${f_{ab}}^{c}$ are the structure constants of a compact semi-simple 
Lie group $G$
\footnotemark[2]$^{)}$.
\footnotetext[2]{$^{)}$ The indices 
$a$, $b$ and $c$ take the values  1, 2, 3, ..., dim$G$.} 
The generators $T_{a}(\sigma)$ also satisfy the hermiticity conditions  
$T_{a}(\sigma)^{\dagger}=T_{a}(\sigma)$. 
(Usually, $\hat{G}_{0}$ is called the loop group of $G$.)  
To take $\hat{G}_{0}$ as the gauge group is essential to derive ^^ ^^ local"  
interactions among local fields.

Let ${\cal A}_{\mu\sigma}^{Y}[x]$ be a Yang--Mills field written as 
\begin{eqnarray}
{\cal A}_{\mu\sigma}^{Y}[x]=\int_{0}^{2\pi}{{d\rho}\over{2\pi}}
{{\cal A}_{\mu\sigma}}^{a\rho}[x]T_{a}(\rho) 
\end{eqnarray}
%
with vector fields ${{\cal A}_{\mu\sigma}}^{a\rho}$ on $\Omega M^{D}$. 
The infinitesimal gauge transformation of ${\cal A}_{\mu\sigma}^{Y}$  
is defined by 
\begin{eqnarray}
\delta{\cal A}_{\mu\sigma}^{Y}[x]=
\partial_{\mu\sigma} \Lambda^{Y}[x]
+i[^{\,}{\cal A}_{\mu\sigma}^{Y}[x], \, \Lambda^{Y}[x]^{\,} ] \;,  
\end{eqnarray}
%
where $\Lambda^{Y}$ is given by 
\begin{eqnarray}
\Lambda^{Y}[x]=\int^{2\pi}_{0} {{d\sigma}\over{2\pi}} 
\Lambda^{a\sigma}[x]T_{a}(\sigma) 
\end{eqnarray}
%
with infinitesimal scalar functions $\Lambda^{a\sigma}$ on $\Omega M^{D}$. 
The reparametrization-invariant condition for $\Lambda^{Y}$ 
is found to be 
\begin{eqnarray}
x'^{\mu}(\sigma) \partial_{\mu\sigma} \Lambda^{Y}[x]
={{\partial\Lambda^{a\sigma}[x]}\over{\partial\sigma}}\,T_{a}(\sigma)  \;, 
\end{eqnarray}
%
which is different from (2). In this case, we conclude that 
$x'^{\mu}(\sigma){\cal A}_{\mu\sigma}^{Y}[x]\neq0$.

The simplest solution of (10) is  
\begin{eqnarray}
\Lambda^{Y(0)}[x]\equiv \int_{0}^{2\pi} {{d\sigma}\over{2\pi}} 
g \lambda^{a}(x(\sigma)) T_{a}(\sigma) \;,
\end{eqnarray}
%
where $g$ is a dimensionless constant and $\lambda^{a}$ are infinitesimal 
scalar functions on $M^{D}$. 
Corresponding to (11), we take   
\begin{eqnarray}
{\cal A}^{Y(0)}_{\mu\sigma}[x]\equiv g {A_{\mu}}^{a}(x(\sigma))  
T_{a}(\sigma) \;
\end{eqnarray}
%
as ${\cal A}^{Y}_{\mu\sigma}$,  
where ${A_{\mu}}^{a}$ are local vector fields on $M^{D}$. 
Substituting (11) and (12) into (8), we have 
the usual gauge transformation of local Yang--Mills fields. 
The field strength of ${\cal A}^{Y(0)}_{\mu\sigma}$ is written in terms of 
that of ${A_{\mu}}^{a}$. 
The Yang--Mills action for ${\cal A}^{Y(0)}_{\mu\sigma}$ reduces to  
the Yang--Mills action for ${A_{\mu}}^{a}$. 
Thus the local Yang--Mills theory can be derived from 
the Yang--Mills theory in loop space.

The next simplest solution of (10) has been considered in ref.[3]. 
We found that the Yang--Mills field ${\cal A}^{Y}_{\mu\sigma}$ associated with 
this solution is written in terms of second-rank tensor fields and 
vector fields on $M^{D}$. 
In this case, the Yang--Mills theory in loop space yields a non-abelian 
Stueckelberg formalism for massive second-rank tensor fields. 

\vspace{1mm}

\section{An extended Yang--Mills theory in loop space}

As a next stage of our discussion, 
let us consider the extension of the Yang--Mills theory in loop space by  
replacing $\hat{G}_{0}$ by its central extension $\hat{G}_{k}$ [4]. 
Hereafter, we refer to the extended theory as the {\it extended} Yang--Mills  
theory (EYMT) in loop space. As will be seen below, the EYMT in loop space  
turns out to be a unified theory of the Yang--Mills and the U(1) gauge  
theories in loop space that are reviewed in the previous section.

The Lie algebra of $\hat{G}_{k}$ is the Kac--Moody algebra 
with the central extension specified by    
\begin{eqnarray}
[^{\,} T_{a}(\rho),\, T_{b}(\sigma)]=
i{f_{ab}}^{c}\delta(\rho-\sigma)T_{c}(\rho)
+ik\kappa_{ab}\delta'(\rho-\sigma)  \;,
\end{eqnarray}
%
where $k$ is a constant called the central charge and $\kappa_{ab}$ is the 
Killing metric of $G$.  
Because of the central extension, the commutator   
$[^{\,}{\cal A}_{\mu\sigma}^{Y}, \, \Lambda^{Y\,}]$ in (8) yields 
terms that do not contain $T_{a}(\rho)$. 
Since the left-hand side of (8), $\delta{\cal A}_{\mu\sigma}^{Y}$, 
is a linear combination of $T_{a}(\rho)$, we have to conclude $k=0$.  
In order to derive non-trivial results,  
we need to modify the Yang--Mills theory in loop space introducing new terms 
into the theory. 
Suitable modification is achieved with the aid of the U(1) gauge theory in  
loop space.

We now add $\Lambda^{U}$ to $\Lambda^{Y}$ and define 
\begin{eqnarray}
\Lambda[x]=\Lambda^{Y}[x]+\Lambda^{U}[x] \;.
\end{eqnarray}
%
The associated gauge field is assumed to be   
\begin{eqnarray}
{\cal A}_{\mu\sigma}[x]={\cal A}_{\mu\sigma}^{Y}[x]
+\widetilde{\cal A}_{\mu\sigma}^{U}[x] \;,
\end{eqnarray}
%
where $\widetilde{\cal A}_{\mu\sigma}^{U}$ is a vector field on 
$\Omega M^{D}$ that does not contain $T_{a}(\rho)$.  
Then the Yang--Mills gauge transformation
\begin{eqnarray}
\delta{\cal A}_{\mu\sigma}[x]=
\partial_{\mu\sigma} \Lambda[x]
+i[^{\,}{\cal A}_{\mu\sigma}[x], \, \Lambda[x]^{\,} ] \;\:  
\end{eqnarray}
%
can be resolved into two parts, that is, the linear combination of 
$T_{a}(\rho)$:  
\begin{eqnarray}
\delta{\cal A}_{\mu\sigma}^{Y}[x]=
\partial_{\mu\sigma} \Lambda^{Y}[x]
+i[^{\,}{\cal A}_{\mu\sigma}^{Y}[x], \, \Lambda^{Y}[x]^{\,} ]^{Y} \;,  
\end{eqnarray}
%
and the other:
\begin{eqnarray}
\delta\widetilde{\cal A}_{\mu\sigma}^{U}[x]=
\partial_{\mu\sigma} \Lambda^{U}[x]
+i[^{\,}{\cal A}_{\mu\sigma}^{Y}[x], \, \Lambda^{Y}[x]^{\,} ]^{U} \;.  
\end{eqnarray}
%
Here $[\;\:\,,\;\:]^{Y}$ denotes the part of a commutator $[\;\:\,,\;\:]$ 
that is written as a linear combination of  $T_{a}(\rho)$, while 
$\,[\;\:\,,\;\:]^{U}$ denotes the other part that occurs owing to the central 
extension. 
The commutation relation (13) is resolved into 
$[^{\,} T_{a}(\rho),\, T_{b}(\sigma)]^{Y}=
i{f_{ab}}^{c}\delta(\rho-\sigma)T_{c}(\rho)$ and  
$[^{\,} T_{a}(\rho),\, T_{b}(\sigma)]^{U}=
ik\kappa_{ab}\delta'(\rho-\sigma)$. 
The gauge transformation (17) is nothing but (8) with (6).  
Comparing (1) with (18), we see that a difference between 
${\cal A}_{\mu\sigma}^{U}$ and $\widetilde{\cal A}_{\mu\sigma}^{U}$ comes 
from the central extension. From (2), (18) and 
$x'^{\mu}(\sigma){\cal A}_{\mu\sigma}^{Y}[x]\neq0$, it follows that 
$x'^{\mu}(\sigma)\widetilde{\cal A}_{\mu\sigma}^{U}\neq0$. 
The vector field $\widetilde{\cal A}_{\mu\sigma}^{U}$ is characterized as the 
gauge field associated with the central extension.

The (naive) field strength of ${\cal A}_{\mu\sigma}$ is 
\begin{eqnarray}
{\cal F}_{\mu\rho,\nu\sigma}  
=\partial_{\mu\rho} {\cal A}_{\nu\sigma}
-\partial_{\nu\sigma} {\cal A}_{\mu\rho} 
+i[^{\,} {\cal A}_{\mu\rho} , \, {\cal A}_{\nu\sigma} ] \;, 
\end{eqnarray}
%
which is a sum of the following two parts:   
\begin{eqnarray}
{\cal F}^{Y}_{\mu\rho,\nu\sigma} \!\!\!&\equiv&\!\!\! 
\partial_{\mu\rho} {\cal A}^{Y}_{\nu\sigma}
-\partial_{\nu\sigma} {\cal A}^{Y}_{\mu\rho} 
+i[^{\,} {\cal A}^{Y}_{\mu\rho} , \, {\cal A}^{Y}_{\nu\sigma} ]^{Y} \;,\quad 
\\
& &\nonumber
\\
{\cal F}^{U}_{\mu\rho,\nu\sigma} \!\!\!&\equiv&\!\!\! 
\partial_{\mu\rho} \widetilde{\cal A}^{U}_{\nu\sigma}
-\partial_{\nu\sigma} \widetilde{\cal A}^{U}_{\mu\rho} 
+i[^{\,} {\cal A}^{Y}_{\mu\rho} , \, {\cal A}^{Y}_{\nu\sigma} ]^{U} \;.\quad
\end{eqnarray}
%
The first part ${\cal F}^{Y}_{\mu\rho,\nu\sigma}$ is the field strength of 
${\cal A}^{Y}_{\mu\sigma}$ and obeys the homogeneous  
gauge-transformation rule, i.e.,   
$\delta{\cal F}_{\mu\rho,\nu\sigma}^{Y}=
i[^{\,}{\cal F}_{\mu\rho,\nu\sigma}^{Y}, \, \Lambda^{Y\,} ]^{Y}$.   
The second part ${\cal F}^{U}_{\mu\rho,\nu\sigma}$ transforms as 
$\delta{\cal F}_{\mu\rho,\nu\sigma}^{U}=
i[^{\,}{\cal F}_{\mu\rho,\nu\sigma}^{Y}, \, \Lambda^{Y\,} ]^{U}$.  
This is an inhomogeneous gauge transformation. 
In order to define a suitable lagrangian for  
$\widetilde{\cal A}^{U}_{\mu\sigma}$, we have  
to find out a ^^ ^^ gauge-invariant" field strength of  
$\widetilde{\cal A}^{U}_{\mu\sigma}$. 
Let us define
\begin{eqnarray}
\widetilde{\cal F}_{\mu\rho,\nu\sigma}^{U}\equiv
{\cal F}^{U}_{\mu\rho,\nu\sigma}
+k\int_{0}^{2\pi}{{d\omega}\over{2\pi}}x'^{\lambda}(\omega)
{\rm Tr}[{\cal A}_{\lambda\omega}^{Y} {\cal F}_{\mu\rho,\nu\sigma}^{Y}] \;,
\end{eqnarray}
%
where ^^ ^^ Tr" denotes the inner product defined by 
 ${\rm Tr}[VW] \equiv \sum_{a,b}\int_{0}^{2\pi} 
{{d\sigma}\over{2\pi}}\kappa_{ab}V^{a\sigma}W^{b\sigma}$. 
(Here $V$ and $W$ are arbitrary elements of the Kac--Moody Lie algebra of 
$\hat{G}_{k}$.) 
Using the condition (10), we can show that   
$\delta\widetilde{\cal F}_{\mu\rho,\nu\sigma}^{U}=0$. 
Hence $\widetilde{\cal F}_{\mu\rho,\nu\sigma}^{U}$ is considered to be the 
field strength of $\widetilde{\cal A}_{\mu\sigma}^{U}$.
The gauge invariance of $\widetilde{\cal F}_{\mu\rho,\nu\sigma}^{U}$ is 
verified for any ${\cal A}_{\mu\sigma}^{Y}$ written as (7), as long as 
$\Lambda^{Y}$ satisfies (10). However, in order that the second term in the 
right-hand side of (22) behaves the same as ${\cal F}_{\mu\rho,\nu\sigma}^{U}$ 
under reparametrizations $\sigma\rightarrow\bar{\sigma}(\sigma)$,  
it is necessary to restrict 
${{\cal A}_{\mu\sigma}}^{a\rho}$ to the following form: 
\begin{eqnarray}
{{\cal A}_{\mu\sigma}}^{a\rho}[x]
=\delta(\sigma-\rho){{\cal A}_{\mu}}^{a\rho}[x] \;.
\end{eqnarray}
%
Here ${{\cal A}_{\mu}}^{a\rho}$ are fields on $\Omega {M}^{D}$ 
that behave as vector functionals on $M^{D}$.  
The Yang--Mills field ${\cal A}_{\mu\sigma}^{Y(0)}$ is a special case of 
${\cal A}_{\mu\sigma}^{Y}$ with (23).

We now define an action for ${\cal A}_{\mu\sigma}^{Y}$ and  
$\widetilde{\cal A}_{\mu\sigma}^{U}$. As done in refs.[1-3], 
we insert the damping factor ${\rm exp}(-L/l^{2})$ with $L\equiv
-\int_{0}^{2\pi} {{d\sigma}\over{2\pi}} \eta_{\mu\nu}x'^{\mu}(\sigma) 
x'^{\nu}(\sigma)$ into the action so that it becomes well-defined. 
Here $l\:(>0)$ is a constant with dimensions of length that gives the size  
of long loops, and $\eta_{\mu\nu}$, ${\rm diag}\eta_{\mu\nu} 
=(1, \,-1, \,-1, \, ..., \,-1)$, is the metric tensor on $M^{D}$. 
The action for ${\cal A}^{Y}_{\mu\sigma}$ and  
$\widetilde{\cal A}^{U}_{\mu\sigma}$ with the damping factor is given by
\begin{eqnarray}
S_{\rm R}={1\over{V_{\rm R}}} \int [dx] ({\cal L}^{Y}+{\cal L}^{U})
{\rm exp}(-{L\over{l^{2}}})   \;  
\end{eqnarray}
%
with the lagrangians
\begin{eqnarray}
{\cal L}^{Y}\!\!\!&=&\!\!\!-{1\over4} k^{Y} 
{\cal G}^{\kappa\rho,\lambda\sigma} {\cal G}^{\mu\chi,\nu\omega} 
{\rm TR}[{\cal F}^{Y}_{\kappa\rho,\mu\chi} 
{\cal F}^{Y}_{\lambda\sigma,\nu\omega}] \;, \:\footnotemark[3]^{)}
\\
& & \nonumber
\\
{\cal L}^{U}\!\!\!&=&\!\!\!-{1\over4} k^{U}
{\cal G}^{\kappa\rho,\lambda\sigma} {\cal G}^{\mu\chi,\nu\omega} 
\widetilde{\cal F}^{U}_{\kappa\rho,\mu\chi} 
\widetilde{\cal F}^{U}_{\lambda\sigma,\nu\omega} \;,
\end{eqnarray} 
%
\footnotetext[3]{$^{)}$ We employ Einstein's convention for indices in the  
loop space  $\Omega M^{D}$; for example,  
$V^{\mu\sigma}W_{\mu\sigma} = \sum_{\mu=0}^{D-1}
\int_{0}^{2\pi} {{d\sigma}\over{2\pi}} V^{\mu\sigma}W_{\mu\sigma}$ .}where  
$[dx]\equiv \prod_{\mu=0}^{D-1} \prod_{n=-\infty}^{\infty} dx^{\mu n}$ 
($x^{\mu n}$ are coefficients of the Fourier expansion 
$x^{\mu}(\sigma)=\sum_{n=-\infty}^{\infty}x^{\mu n}e^{in\sigma})$, 
$V_{\rm R} \equiv \int \{ dx \} {\rm exp}(-L/l^{2})$  
with  $ \{ dx \} \equiv \prod_{\mu=0}^{D-1} \prod_{n=-\infty}
^{\infty, \, n\neq0} dx^{\mu n}$, and $k^{Y}$ and $k^{U}$ are constants. 
The (inverse) metric tensor ${\cal G}^{\mu\rho,\nu\sigma}$ and  
the inner product ^^ ^^ TR", which were defined in ref.[1] and ref.[3] 
respectively, are necessary to guarantee reparametrization invariance of 
${\cal L}^{Y}$ and ${\cal L}^{U}$.  
In this paper, however, we use the metric 
$\eta^{\mu\nu}\delta(\rho-\sigma)$ as a form of ${\cal G}^{\mu\rho,\nu\sigma}$ 
in a certain gauge of reparametrizations. Similarly, we use the inner product 
Tr as a form of TR.  
As a result, ${\cal L}^{Y}$ and ${\cal L}^{U}$ are no longer  
reparametrization-invariant, while they are still gauge-invariant.

\vspace{1mm}

\section{The Chapline--Manton coupling}

In this section we derive a local field theory describing coupling of  
${A_{\mu}}^{a}$ and $B_{\mu\nu}$ from the EYMT in loop space. 
Substituting (11) and (12) into (17), we have 
the usual gauge transformation of local Yang--Mills fields: 
\begin{eqnarray}
\delta {A_{\mu}}^{a}(x)=
\partial_{\mu} \lambda^{a}(x)-g {A_{\mu}}^{b}(x) \lambda^{c}(x) 
{f_{bc}}^{a} \;. 
\end{eqnarray}
%
Similarly, substitution of (11) and (12) into (20) yields  
\begin{eqnarray}
{\cal F}^{Y(0)}_{\mu\rho,\nu\sigma}[x]
=g \delta(\rho-\sigma){F_{\mu\nu}}^{a}
(x(\sigma)) T_{a}(\sigma) \;\,
\end{eqnarray}
%
with
${F_{\mu\nu}}^{a} \equiv 
\partial_{\mu}{A_{\nu}}^{a}-\partial_{\nu}{A_{\mu}}^{a}
-g{A_{\mu}}^{b}{A_{\nu}}^{c}{f_{bc}}^{a}$.
They were already given in the Yang--Mills theory in loop space [3].

Since $\widetilde{\cal A}_{\mu\sigma}^{U}$ dose not satisfy (3),  
we can not take ${\cal A}_{\mu\sigma}^{U(0)}$ as 
$\widetilde{\cal A}_{\mu\sigma}^{U}$. Thus, instead of 
${\cal A}_{\mu\sigma}^{U(0)}$, we consider 
\begin{eqnarray}
\widetilde{\cal A}_{\mu\sigma}^{U(0)}[x]\equiv
qx'^{\nu}(\sigma)\{ B_{\mu\nu}(x(\sigma))+C_{\mu\nu}(x(\sigma)) \} \;,
\end{eqnarray}
%
introducing the local symmetric tensor field $C_{\mu\nu}$ on $M^{D}$. 
Obviously $\widetilde{\cal A}_{\mu\sigma}^{U(0)}$ does not satisfy (3). 
Substituting (4), (11), (12) and (29) into (18), we obtain 
\begin{eqnarray}
\delta B_{\mu\nu}(x) \!\!\!&=&\!\!\! \partial_{\mu}\lambda_{\nu}(x)
-\partial_{\nu}\lambda_{\mu}(x)
-\tilde{k}{A_{[\mu}}^{a}(x)\partial_{\nu]}\lambda_{a}(x) \;, \quad \: \\
& & \nonumber \\
\delta C_{\mu\nu}(x) \!\!\!&=&\!\!\! 
-\tilde{k}{A_{(\mu}}^{a}(x)\partial_{\nu)}\lambda_{a}(x) \;,
\end{eqnarray}
%
where $\tilde{k}\equiv kg^{2}/2q$. 
The lowering of the index $a$ has been done with $\kappa_{ab}$. 
Notice that the Yang--Mills fields ${A_{\mu}}^{a}$ occur in the gauge 
transformations of $B_{\mu\nu}$ and $C_{\mu\nu}$. 
In terms of the infinitesimal vector parameter $\xi_{\mu}\equiv\lambda_{\mu}+
\tilde{k}\lambda_{a}{A_{\mu}}^{a}$, (30) is rewritten as 
\begin{eqnarray}
\delta B_{\mu\nu}(x)=\partial_{\mu}\xi_{\nu}(x)
-\partial_{\nu}\xi_{\mu}(x)
-\tilde{k}\lambda_{a}(x)\partial_{[\mu}{A_{\nu]}}^{a}(x) \;.
\end{eqnarray}
%
Similarly, defining the tensor field 
$\widetilde{C}_{\mu\nu}\equiv C_{\mu\nu}+\tilde{k}{A_{\mu}}^{a}A_{\nu a}$, 
we can simply write (31) as $\delta \widetilde{C}_{\mu\nu}(x)=0$, 
from which we see that $\widetilde{C}_{\mu\nu}$ is gauge-invariant.

Substitution of (12), (28) and (29) into (22) leads to  
\begin{eqnarray}
\widetilde{\cal F}_{\mu\rho,\nu\sigma}^{U(0)}[x]
\!\!\!&=&\!\!\! 
q\delta(\rho-\sigma)x'^{\lambda}(\sigma)
\{ H_{\lambda\mu\nu}(x(\sigma))
+\partial_{[\mu}\widetilde{C}_{\nu]\lambda}(x(\sigma)) \} 
\nonumber \\
& &\!\!\!-q\delta'(\rho-\sigma)
\{ \widetilde{C}_{\mu\nu}(x(\rho))+\widetilde{C}_{\mu\nu}(x(\sigma)) \} \;,
\end{eqnarray}
%
where 
\begin{eqnarray}
H_{\lambda\mu\nu}(x)
\equiv F_{\lambda\mu\nu}(x)+\tilde{k}\Omega_{\lambda\mu\nu}(x) 
\end{eqnarray}
%
with 
\begin{eqnarray}
F_{\lambda\mu\nu} \!\!\!&\equiv&\!\!\! \partial_{\lambda}B_{\mu\nu}
+\partial_{\mu}B_{\nu\lambda}+\partial_{\nu}B_{\lambda\mu} \;, \\
& & \nonumber \\ 
\Omega_{\lambda\mu\nu} \!\!\!&\equiv&\!\!\!
{A_{[\lambda}}^{a}\partial_{\mu}A_{\nu]a}
-{g\over3}f_{abc} {A_{[\lambda}}^{a} {A_{\mu}}^{b} {A_{\nu]}}^{c} \;.
\end{eqnarray}
%
The totally antisymmetric tensor $\Omega_{\lambda\mu\nu}$ is nothing other 
than the Chern--Simons 3-form. As is easily shown, $H_{\lambda\mu\nu}$ is  
invariant under (27) and (30) (and under (27) and (32)). Consequently, 
$\widetilde{\cal F}_{\mu\rho,\nu\sigma}^{U(0)}$ is gauge-invariant as  
we expected. The fact that $\widetilde{\cal F}_{\mu\rho,\nu\sigma}^{U(0)}$ has 
been obtained in a gauge-invariant form justifies (22), the definition of 
$\widetilde{\cal F}_{\mu\rho,\nu\sigma}^{U}$.  
The gauge transformation (32) and the field strength $H_{\lambda\mu\nu}$ 
were first found by Chapline and Manton in a heuristic manner in the study of  
a unification of supergravity and supersymmetric Yang--Mills theory [5].

Finally, let us discuss how the action $S_{\rm R}$ is written in terms of 
the local fields [1-3]. 
We substitute (28) and (33) into (25) and (26), respectively.  
Being carried out integrations with respect to $\rho$, $\chi$ and $\omega$,  
the lagrangians ${\cal L}^{Y}$ and ${\cal L}^{U}$ take forms of integral 
with respect to $\sigma$. In the action $S_{\rm R}$ with these lagrangians,  
we expand the functions of $x^{\mu}(\sigma)$,  
such as $H_{\lambda\mu\nu}(x(\sigma)){H_{\kappa}}^{\mu\nu}(x(\sigma))$,  
around $(x^{\mu0})$. 
Then, all the differential coefficients at  
$(x^{\mu0})$ in each Taylor series vanish after integration with  
respect to $x^{\mu0}$, since $-\infty<x^{\mu0}<\infty$. As a result, 
we have the action in which the arguments 
$x^{\mu}(\sigma)$ of the functions are replaced by $x^{\mu0}$.     
Carrying out the integrations with respect to $x^{\mu n}$ $(n\neq0)$ 
finally, \footnotemark[4]$^{)}$ 
\footnotetext[4]{$^{)}$  The integrations with respect to $x^{0n}$ are carried 
out after applying the Wick rotation $x^{0n}\rightarrow-ix^{0n}$.}  
we obtain 
\begin{eqnarray}
S_{\rm R}^{(0)}\!\!\!&=&\!\!\! \int d^{D}x
[-{1\over4}F_{\mu\nu a}(x)F^{\mu\nu a}(x)
+{1\over12}H_{\lambda\mu\nu}(x)H^{\lambda\mu\nu}(x) \nonumber \\
& &+{1\over12}\partial_{[\mu}\widetilde{C}_{\nu]\lambda}(x)
\partial^{[\mu}\widetilde{C}^{\nu]\lambda}(x)
-{1\over6}m^{2}\widetilde{C}_{\mu\nu}(x)\widetilde{C}^{\mu\nu}(x)] \;,
\end{eqnarray}
%
where $x^{\mu0}$ have been replaced by $x^{\mu}$. 
In deriving this action, we have set the normalization conditions 
$3k^{U}q^{2}l^{2}\delta(0)^{2}/2=1$ and $k^{Y}g^{2}\delta(0)^{2}=1$ and 
have defined $m^{2}$ by $m^{2}\equiv -6k^{U}q^{2}\delta''(0)$,  
all of which expressions are understood with suitable regularizations of the  
$\delta$-functions. 
As seen from (37), the Yang--Mills fields ${A_{\mu}}^{a}$ and  
the antisymmetric tensor field $B_{\mu\nu}$ are massless, while 
the symmetric tensor field $\widetilde{C}_{\mu\nu}$ has mass $m$ that is  
inversely proportional to the loop-size $l$.  
In addition, we see that $B_{\mu\nu}$ is coupled to ${A_{\mu}}^{a}$ via the 
Chern--Simons 3-form $\Omega_{\lambda\mu\nu}$, while $\widetilde{C}_{\mu\nu}$ 
is free from ${A_{\mu}}^{a}$ and $B_{\mu\nu}$.

\vspace{1mm}

\section{Discussion}

A unification of the Yang--Mills and the U(1) gauge theories in loop space  
has been achieved with the EYMT in loop space. 
From this theory, we have derived the gauge transformation (30) 
and the field strength $H_{\lambda\mu\nu}$ in a systematic manner; some of the 
results by Chapline and Manton are obtained within the framework of 
Yang--MIlls theories. 
In ref.[5], they found the ^^ ^^ finite" gauge transformation 
of $B_{\mu\nu}$ by utilizing Chern's formula. If its gauge parameter is 
unlimitedly near the identity element, the finite gauge transformation reduces 
to the third term in the rigth-hand side of (30). 
It is important to derive the finite gauge transformation 
from the EYMT in loop space in order to complete our discussion in the present 
paper.

It is known that $B_{\mu\nu}$ obeying the gauge transformation (30)(or (32)) 
plays an essential role in anomaly cancellations in field theories containing 
chiral fermions. A well-known example in which anomaly cancellations due to 
$B_{\mu\nu}$ are automatically incorporated is the type I superstring theory 
based on the gauge group SO(32) [6]. 
Since the EYMT in loop space naturally yields (30), 
we might be able to understand the anomaly-cancellation mechanism in terms  
of the EYMT in loop space.

From the EYMT in loop space, we can derive a local field theory that describes 
couplings of non-abelian second-rank tensor fields and abelian third-rank 
tensor fields. It is done by employing the nonlinear realization of the 
Kac--Moody gauge group $\hat{G}_{k}$. 
The details will be reported in the forthcoming paper.

\vspace{1.5cm}

\noindent
{\Large\bf Acknowledgments}

\vspace{5mm}

We are grateful to Professor S. Naka and other members of the Theoretical  
Physics Group at Nihon University for their encouragements and useful commen
ts .We would like to thank Professor T. Fujita for his careful reading of the 
manuscript. 
This work was supported by Nihon University Research Grant for 1995.

\newpage

\begin{center}
{\Large\bf References}
\vspace{1mm}

\end{center}
\begin{enumerate}
%
\item S. Deguchi and T. Nakajima, Int. J. Mod. Phys. {\bf A9} (1994) 1889. 
\item S. Deguchi and T. Nakajima, Prog. Theor. Phys. {\bf 94} (1995) 305. 
\item S. Deguchi and T. Nakajima, Int. J. Mod. Phys. {\bf A10} (1995) 1019. 
\item A. Pressley and G. Segal, Loop groups (Oxford Univ. Press, New York, 
1986) 

\vspace{1mm}

B. E. Baaquie, Nucl Phys {\bf B295} (1988) 188.
\item G. F. Chapline and N. S. Manton, Phys. Lett. {\bf B120} (1983) 105. 
\item M. B. Green and J. H. Schwarz, Phys. Lett. {\bf B149} (1984) 117; 
{\bf B151} (1985) 21. 

\vspace{1mm}

M. B. Green, J. H. Schwarz and P. C. West, Nucl. Phys. {\bf B254}, (1985) 327.  

\end{enumerate}

\end{document}